\documentclass[reprint,amsmath,amssymb,longbibliography,aps, prb]{revtex4-2}
\pdfoutput=1
\usepackage{hyperref}
\usepackage{url}
\usepackage{color}
\usepackage{acro}
\usepackage{chemmacros}
\usepackage{svg}
\usepackage[normalem]{ulem}

\usepackage{graphicx}

\begin{document}
\title{Topological superconductivity driven by correlations and linear defects in multiband superconductors}

\author{Mainak Pal$^{1*}$\thanks{email for correspondence: mainak.pal@ufl.edu}, Andreas Kreisel$^{2,3}$, and P.J. Hirschfeld$^1$}

\affiliation{$^1$Department of Physics, University of Florida, Gainesville, Florida, USA}
\affiliation{$^2$Institut f\" ur Theoretische Physik, Universit\"at Leipzig, D-04103 Leipzig, Germany}
\affiliation{$^3$Niels Bohr Institute, University of Copenhagen, 2100 Copenhagen, Denmark}

\begin{abstract}
There have been several proposals for platforms sustaining topological superconductivity in high temperature superconductors, in order to make use of the larger superconducting gap and the expected robustness of Majorana zero modes towards perturbations. In particular, the iron-based materials offer relatively large $T_c$ and nodeless energy gaps. In addition, atomically flat surfaces enable the engineering of
defect structures and the subsequent measurement of spectroscopic properties to reveal topological aspects.
From a theory perspective, a materials-specific description is challenging due to the correlated nature of the materials and complications arising from the multiband nature of the electronic structure.
Here we include both aspects in realistic interacting models, and find that the correlations themselves  can lead to local magnetic order close to linear potential scattering defects at the surface of the superconductor. Using a self-consistent Bogoliubov-de Gennes framework in a real-space setup using a prototype electronic structure, we allow for arbitrary magnetic orders and show how a topological superconducting state emerges. The calculation of the topological invariant and the topological gap allows us to map out the phase diagram for the case of a linear chain of potential scatterers. While intrinsic spin-orbit coupling is not needed to enter the topological state in presence of  spin-spiral states,  it enlarges the topological phase. We discuss the interplay of a triplet component of the superconducting order parameter and the spin spiral leading effectively to extended spin orbit coupling terms, and connect our results to experimental efforts on the Fe(Se,Te) system. 
 \end{abstract}

\maketitle

\section{Introduction}
Signatures of nontrivial topology in condensed matter systems 
have been a holy grail at the frontiers of unconventional superconductivity for quite some time now, more so because of its promising role in creation of protected zero-energy modes that can find application in robust quantum computation in a fault-tolerant way \cite{RevModPhys.80.1083}.
The prototype system where zero-energy modes appear is the Kitaev chain \cite{Kitaev_2001}, a one-dimensional chain with spinless Fermions exhibiting $p$-wave superconductivity and thus inducing Majorana zero modes (MZMs) at the ends of the chain. These quantum states are protected by symmetry, do not hybridize due to their exponential localization and therefore stay at zero energy.

Realizations of topological superconductivity remain elusive since real electrons carry spin degrees of freedom, and $p$-wave superconductivity, i.e. triplet pairing is rare in nature. Common proposals for possible platforms that sustain an (effectively) topological character are based on proximitized superconductivity on nanowires with strong spin orbit coupling \cite{Lutchyn10,Oreg10,Mourik12,Lutchyn18} or by introducing magnetic structures on top of conventional superconducting substrates
\cite{PhysRevLett.100.096407, PhysRevLett.104.040502,Nad13}
For example, placing a chain of ferromagnetic metal atoms on a superconductor yields hybridized Shiba bound states and allows one to tune into the topological regime in the presence 
of Rashba spin-orbit coupling \cite{Pientka13, PhysRevB.90.235433,Brydon15} and {was} subsequently
tested in  experiment
\cite{Nadj-Perge602,Ruby15,Paw16,Kim18}. 

From an engineering point of view, the platforms based on nanowires or atomic chains on (conventional) s-wave superconductors like Al or Pb are easier to fabricate and control experimentally, compared to unconventional superconductors like the Fe-based material Fe(Se,Te). On the other hand, unconventional superconductors have relatively higher critical superconducting temperature and robust gaps while hosting  significant spin-orbit interaction which are favorable characteristics for achieving the topological state.

Recently, apparent Majorana zero-modes at the ends of naturally occurring atomic line defects on the surface of Fe-based materials were reported in experiments~\cite{Chen2020,doi:10.1126/science.aaw8419,Li2022}.
In {a} scanning-tunneling microscopy (STM) study, the authors identified line defects on the surface of monolayer Fe(Te,Se) arising from missing Te/Se atoms in the surface layer of Se/Te~\cite{Chen2020}. Sharp spectral  peaks in the conductance at zero energy were observed  only at the ends of these chains.
In this  discovery  paper,  the 
missing line of Se/Te atoms was hypothesized to have a twofold effect:
(a) inducing a change in the local chemical potential because of the missing Se/Te charges, and
(b) creating a local Rashba type interaction via
i) inversion symmetry breaking due to removal of Se/Te atoms from only one sublayer and
ii) local electrostatic fields due to missing charges.

However, these authors did not rule out the possibility that removal of the Se/Te atoms can expose the magnetic nature of corresponding Fe atoms, thus mimicking presence of an effective magnetic chain, as suggested in the magnetic adatom proposal \cite{PhysRevB.90.235433,Nad13}.  
In one theoretical scenario, it was reported that first principle calculations show large
density of states near the Fermi level due to missing Te/Se atoms that induces magnetic order on the line defects \cite{wu2020topological}. The authors then used a
single band model {with phenomenological exchange interactions} to suggest that in either the ferromagnetic or or antiferromagnetic (AFM) configuration of the chain, it undergoes a topological transition and hosts MZMs. The work in Ref.~\cite{PhysRevX.11.011041} employs a single band model description  and relies on a significant Rashba spin-orbit interaction to turn the chain into a topological superconductor. However, it is not a priori clear whether breaking of inversion symmetry, which induces Rashba spin-orbit coupling on the surface, is a necessary ingredient.

While intriguing by themselves, these studies propose different phenomenological pictures and mechanisms, indicating that the issue is far from settled theoretically. Further investigations are required to {deduce the correct effective low-energy model describing the topological transition, as well as to understand how it arises from realistic descriptions of Fe-based systems.} In our work, we employ a multiorbital electronic structure model including local Hubbard-Kanamori interactions relevant for Fe-based systems in order  to understand the emergence of the topological phase.  We  treat superconducting order and magnetic order emerging from the same band structure { on the same footing} in a self-consistent mean field framework.
Our approach is capable of generating magnetic order close to the defect structures which are a priori most likely to act primarily by inducing potential scattering, i.e. a local shift of the chemical potential. More concretely, we model our Hamiltonian to describe a linear chain of potential scatterers. In a previous work we used a similar model to describe local magnetic order causing magnetic anisotropy in presence of spin-orbit coupling and Hubbard-Kanamori type correlations~\cite{PhysRevB.103.245132}.

Here, we show that in realistic band-structures of iron-based superconductors   in the vicinity of a magnetic instability,
linear potential defects can drive {noncollinear} local magnetic states. These in turn modify the nature of the superconducting order parameter and therefore { drive the system into the topological phase.}
This effect can be understood from the viewpoint that already  in the homogeneous system (without { the defect}), the spin-orbit coupling induces small components of (pseudospin) triplet pairing, once unconventional  pairing is present. To detect the topological phase in our approach, we calculate the topological invariant using based Pfaffians on time-reversal invariant momenta (TRIM) as constructed from the real space Hamiltonian using a supercell method (see {Appendix \ref{supercell_method}}). Since the bulk superconductor exhibits trivial topology, we then expect MZMs on the endpoints of chains of potential scatterers.

\begin{figure}[tb]
\centering
\includegraphics[width=1\linewidth]{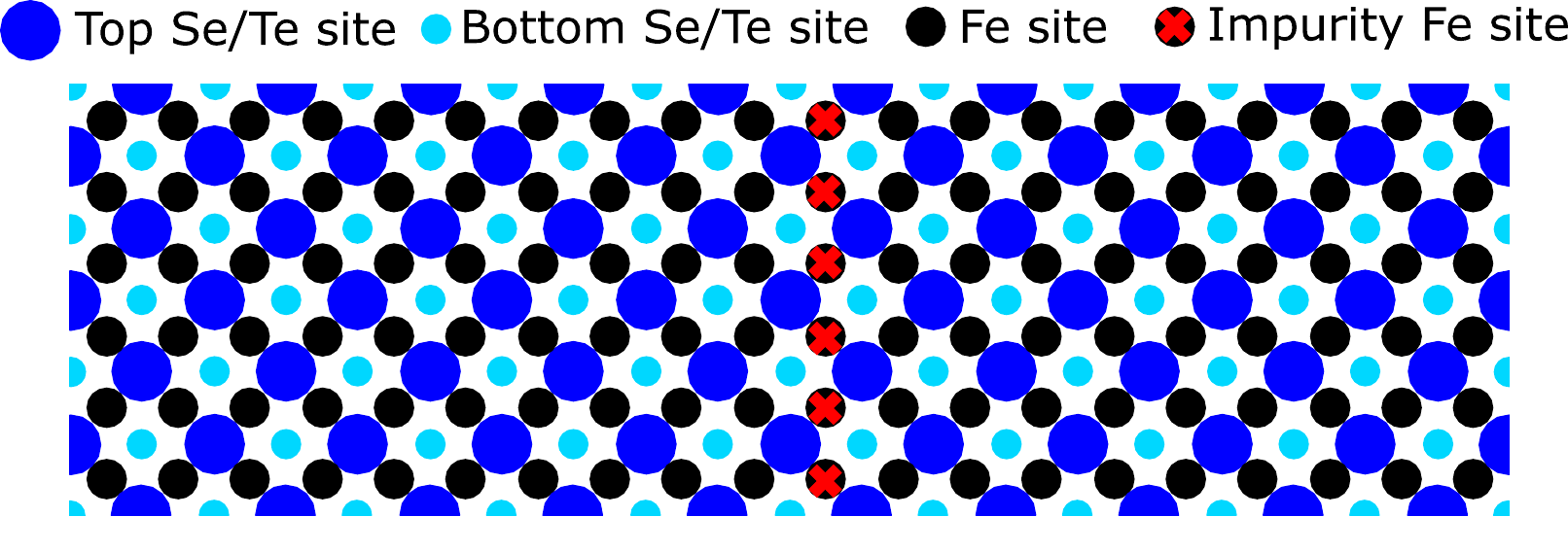}
\caption[A schematic ``vertical" chain of potential impurity inserted in a $20\ {\times}\ 6$ system]{A schematic ``vertical" chain of potential impurity inserted in a $20\ {\times}\ 6$ system that we work with. Strength of the potential impurity can be tuned at every site. Periodic boundary condition is imposed in both horizontal and vertical directions, thereby effectively simulating a one-dimensional infinite impurity or rather an impurity circle on the surface of a toroid.
{}
}
\label{fig:system_with_impurity}
\end{figure}

\section{Model}

Our theoretical calculations can be summarized in three conceptionally different steps as follows. First, we consider a homogeneous system and calculate the superconducting order parameter and magnetism in  a self-consitent mean-field approach to (i) characterize the superconducting order parameter and its modifications from the usual picture once spin-orbit coupling is present and (ii) tune the parameters of the Hubbard-Kanamori interactions close to, but below the magnetic instability (while not considering superconductivity).

Second, we introduce non-magnetic impurities (red crosses) into our description in real space and thereby use a system size of $20 \times 6$ lattice points, see Fig.~\ref{fig:system_with_impurity}, to again calculate self-consistently the spatial pattern of superconducting order parameter and (local) magnetic moments.  We employ periodic boundary conditions in $x$ and $y$ directions for tight-binding hoppings and superconducting order parameters. 

Finally, we construct the Hamiltonian of a system that consists of repeating units in one dimension of the Hamiltonian treated in mean field approximation, and examine the resulting bands  as function of the supercell momentum (as quantum number) for the periodically repeating units.  This allows us to  calculate the topological invariant associated with such a quasi-one dimensional superconductor.
\begin{figure*}[tb]
    \centering
    \includegraphics[width=1\linewidth]{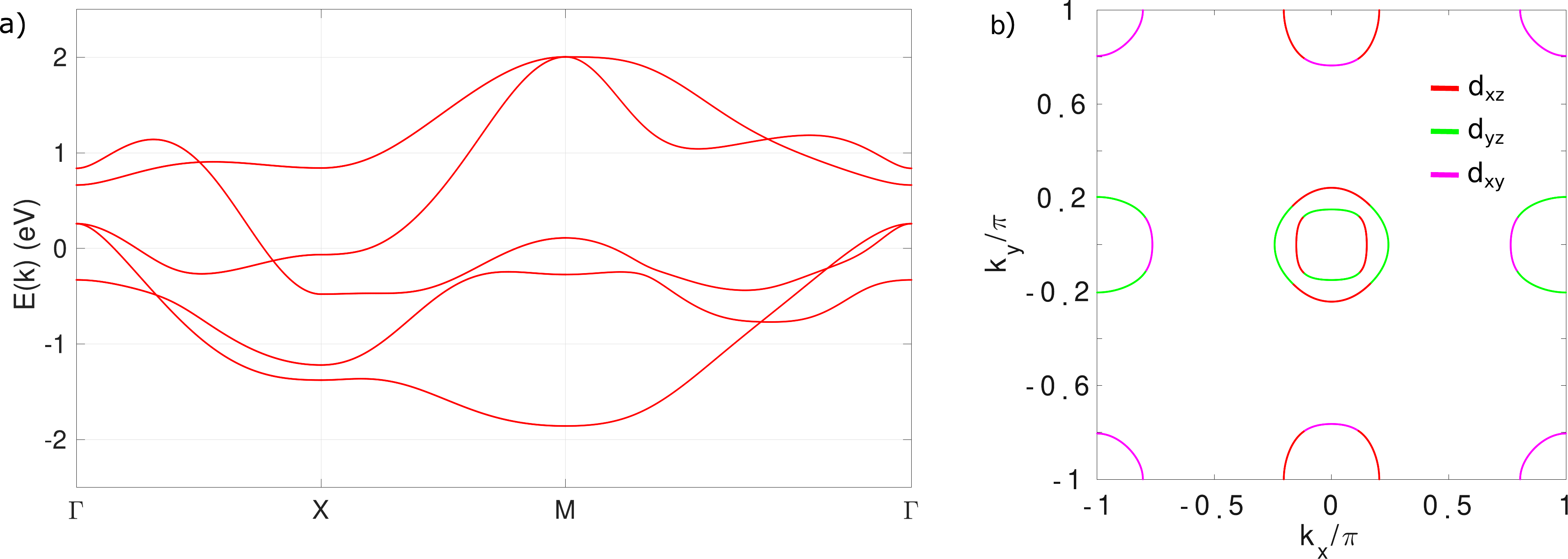}
    \caption{{Band structure and Fermi surface of the system:} (a) Band structure of the system along the high-symmetry path in the Brillouin zone and (b) Fermi surface along with the dominant orbital character of the bands at a filling of $n=5.95$. 
    }
    \label{fig:combined_orbital_weights_calc_files_U0_Vimp0_SOC_0pt02_for_Fermi_Surf_bandstructure_inkscape}
\end{figure*}

All three steps use the following Hamiltonian which consits of 5 distinct terms,
\begin{align}
H = H_{0} + H_{\mathrm{int}}  + H_{\mathrm{imp}} + H_{\mathrm{SC}} + H_{\mathrm{SOC}}
\label{eq:all}
\end{align}
which we introduce in detail in the following sub-sections. Note that first step does not include the impurity term, $ H_{\mathrm{imp}}$ and superconducting term $H_{\mathrm{SC}}$ when we discuss  the magnetic phase diagram of the normal state.
\subsection{Tight-binding Hamiltonian}
In order to capture essential properties of the electronic structure of Fe-based materials, we employ, for the normal state Hamiltonian, a tight-binding model~\cite{PhysRevB.81.054502} in two dimensions with five orbital states on each lattice point,
\begin{equation}
H_\mathrm{0}  = {\displaystyle \sum_{
 ij\mu\nu\sigma}\Big(t^{\mu\nu}_{ij}c_{i\mu\sigma} - \delta_{ij}\delta_{\mu\nu}\mu_0\Big)c^{\dagger}_{i\mu\sigma}c_{j\nu\sigma}}\,.
\label{eq:H0}
\end{equation}
where $c^{\dagger}_{i\mu\sigma}$ creates an electron on lattice point $i$ and orbital $\mu$ in spin state $\sigma$. For all self-consistent calculations of a setup of the Hamiltonian in real space, the chemical potential $\mu_0$ is determined in order to reach the filling $n$.
Indeed, the true crystal structure of Fe-based materials contains two inequivalent iron atoms such that  10 orbital states per elementary cell would be needed. However, the Fe atoms are related by glide plane symmetry (which consists of a reflection on the Fe-lattice plane followed by a translation connecting neighbored Fe atoms such that the (non-relativistic) electronic structure can be exactly downfolded to a 5 orbital model. Deviations from this are (i) pertubartive in the spin-orbit coupling which is small for the 11, 111 and 1111 Fe-based systems~\cite{PhysRevB.80.104503,PhysRevLett.114.107002, Borisenko2016} and (ii) become qualitative for 122 systems once coupling in the third dimension is considered.

\subsection{Hubbard-Kanamori Repulsion}

To capture effects of correlations, we start with a local Hubbard-Kanamori interaction \cite{10.1143/PTP.30.275,PhysRevB.99.014509,doi:10.1146/annurev-conmatphys-020911-125045},

\begin{equation}
\begin{split}
H_{\mathrm{int}} & = U \displaystyle \sum_{i\mu} n_{i\mu\uparrow} n_{i\mu\downarrow} + U^{\prime}\displaystyle\sum_{i,\mu<\nu,\sigma}n_{i\mu\sigma}n_{i\nu\sigma^{\prime}}
\\&+ \big(U^{\prime} - J\big)\displaystyle\sum_{i,\mu<\nu,\sigma}n_{i\mu\sigma}n_{i\nu\sigma}
+J^{\prime}\displaystyle\sum_{i,\mu\neq\nu}c_{i\mu\uparrow}^{\dagger}c_{i\mu\downarrow}^{\dagger}c_{i\nu\downarrow}c_{i\nu\uparrow}
\\&
+J\displaystyle\sum_{i,\mu<\nu,\sigma}c_{i\mu\sigma}^{\dagger}c_{i\nu\sigma^{\prime}}^{\dagger}c_{i\mu\sigma^{\prime}}c_{i\nu\sigma}
\label{eq:H_HbHu}
\end{split}
\end{equation}
where $\sigma$ and $\sigma^{\prime}$
are {complementary} spin indices, $i$ is the site for the local interaction and $\mu, \nu$ are $d$ orbitals.
The Hubbard repulsion term is parametrized by $U$ and involves opposite spins in the same orbital, while the second term is its generalization to multiple orbitals where occupation of states with opposite spins in different orbitals is penalized with the energy $U'$.
The third term with coefficient $U^{\prime} - J$ describes interaction between electrons with parallel spin in different orbitals, and thus enforces Hund's rules.
The fourth term $J'$ allows for hopping of spin-pairs between orbitals. We ignore orbital differentiations of interaction parameters and set $J = J^{\prime}$, $U = U' + 2J$, which correspond to spin-rotationally invariant interactions. 
$J = J^{\prime}$ is guaranteed by choosing a basis of real-valued Wannier states~\cite{doi:10.1146/annurev-conmatphys-020911-125045} and we work with $J = J' = U/4$ in the following, a reasonable choice for Fe-based materials.

The interaction Hamiltonian has been decoupled at the mean-field level in this work such that the interaction Hamiltonian is given by \cite{PhysRevB.99.014509}
\begin{equation}
    \begin{split}
     &H_{\mathrm{int}}^{\mathrm{MF}}  \ = \displaystyle\sum_{i\nu\sigma}\big[U\langle n_{i\nu\sigma^{\prime}}\rangle + \displaystyle\sum_{\mu\ne\nu}\big\{U^{\prime}\langle n_{i\mu\sigma^{\prime}}\rangle 
       + \big(U^{\prime}-J\big)\langle n_{i\mu\sigma} \rangle \big\}\big]\\&\ \ \ \ \ \ \ \ \ \ \ \ \ \ \ \times c_{i\nu\sigma}^{\dagger}c_{i\nu\sigma} \\
      &- \displaystyle\sum_{i,\mu\ne\nu,\sigma}\big[\big(U^{\prime} -J \big)\langle c_{i\nu\sigma}^{\dagger}c_{i\mu\sigma} \rangle -J^{\prime} \langle c_{i\mu\sigma^{\prime}}^{\dagger}c_{i\nu\sigma^{\prime}}\rangle
        - J\langle c_{i\nu\sigma^{\prime}}^{\dagger}c_{i\mu\sigma^{\prime}}\rangle \big]\\&\ \ \ \ \ \ \ \ \ \ \ \ \ \ \ \times c_{i\mu\sigma}^{\dagger}c_{i\nu\sigma} \\
     &-\displaystyle\sum_{i\nu\sigma}\big[ U\langle c_{i\nu\sigma}^{\dagger}c_{i\nu\sigma^{\prime}}\rangle + J\sum_{\mu\ne\nu}\langle c_{i\mu\sigma}^{\dagger}c_{i\mu\sigma^{\prime}}  \rangle\big]c_{i\nu\sigma^{\prime}}^{\dagger}c_{i\nu\sigma}\\
      &-\sum_{i,\mu\ne\nu,\sigma}\big[U^{\prime}\langle c_{i\nu\sigma}^{\dagger}c_{i\mu\sigma^{\prime}}\rangle + J^{\prime}\langle c_{i\mu\sigma}^{\dagger}c_{i\nu\sigma^{\prime}} \rangle\big]c_{i\mu\sigma^{\prime}}^{\dagger}c_{i\nu\sigma}
     \label{eq:H_HbHu_MF}
    \end{split}
\end{equation}
where $\sigma^{\prime}$ 
is complementary spin index of $\sigma$ and vice versa and the mean fields $\langle \ldots \rangle$ are calculated {as expectation values at temperature $T$} from the eigenenergies and eigenstates of the total Hamiltonian.
Once the expectation values become spin-dependent, there is a finite magnetization at lattice point $i$,
\begin{equation}
\textbf{M}_i = \left(\begin{array}{c}
                      \langle c_{i,\uparrow}^{\dagger}c_{i,\downarrow}+c_{i,\downarrow}^{\dagger}c_{i,\uparrow}\rangle\\
                     - i\langle c_{i,\uparrow}^{\dagger}c_{i,\downarrow}-c_{i,\downarrow}^{\dagger}c_{i,\uparrow}\rangle\\
                      \langle c_{i,\uparrow}^{\dagger}c_{i,\uparrow}-c_{i,\downarrow}^{\dagger}c_{i,\downarrow} \rangle
                     \end{array}\right)
 \label{eq:magnetization}
 \end{equation}
\subsection{Impurity Potential}

We consider a chain of impurities  characterized by  a local non-magnetic scattering potential of  $V_{\mathrm{imp}}$ which is included in the real space Hamiltonian as

\begin{equation}
H_{\mathrm{imp}} = V_{\mathrm{imp}}\sum_{i^{*}\mu\sigma}c_{i^{*}\mu\sigma}^{\dagger}c_{i^{*}\mu\sigma}
\end{equation}
where $i^{*}$ indexes the {set of impurity positions} along  the $y$-axis, see  Fig. \ref{fig:system_with_impurity}. These represent a  line defect in the $20\ {\times}\ 6$ system, on which all results shown in this work are based.

\subsection{Superconducting Hamiltonian}

Superconductivity is described by the pairing term,
\begin{equation}
\begin{split}
\displaystyle
H_{\mathrm{SC}} = \frac{1}{2}\sum_{m=0}^{3}\sum_{\tilde i\tilde j}\sum_{\alpha_{1,2,3,4}}&\Gamma_{m\tilde j\tilde i}\big(\sigma_{m}i\sigma_{2}\big)^{\dagger}_{\alpha_{3}\alpha_{1}}\big(\sigma_{m}i\sigma_{2}\big)_{\alpha_{2}\alpha_{4}} \\
 & \times \big(c_{\tilde i \alpha_{1}}c_{\tilde j\alpha_{3}}\big)^{\dagger}c_{\tilde i \alpha_{2}}c_{\tilde j\alpha_{4}}
\label{eq:H_SC}
\end{split}
\end{equation}
where $\tilde i,\tilde j$ now include both the lattice and orbital indices collectively { e.g. $\tilde i = (i,\nu)$}, and $\Gamma_{\tilde i\tilde j}$ are pairing interactions between 
$\tilde i$ and $\tilde j$ mediating unconventional electron pairing. The spin indices, represented by $\alpha_{1}$, $\alpha_{2}$, $\alpha_{3}$, $\alpha_{4}$ (written in short as $\alpha_{1,2,3,4}$ in Eq. (\ref{eq:H_SC}) to avoid clutter), can each be either $\uparrow$ or $\downarrow$; $m = 0$ denotes the singlet channel and $m = 1,2,3$ denote the triplet channels.

A mean field decoupling yields the term
\begin{equation}
\begin{split}
H_{\mathrm{SC}}^{\mathrm{MF}}  = \frac{1}{2}&\sum_{m=0}^{3}\sum_{\tilde i\tilde j}\sum_{\alpha_{1\sim4}}\Gamma_{m\tilde j\tilde i}\left(\sigma_{m}i\sigma_{2}\right)_{\alpha_{3}\alpha_{1}}^{\dagger}\left(\sigma_{m}i\sigma_{2}\right)_{\alpha_{2}\alpha_{4}}
\\
 &\times \big\{ \langle c_{\tilde j\alpha_{3}}^{\dagger}c_{\tilde i\alpha_{1}}^{\dagger}\rangle c_{\tilde i\alpha_{2}}c_{\tilde j\alpha_{4}} + c_{\tilde j\alpha_{3}}^{\dagger}c_{\tilde i\alpha_{1}}^{\dagger}\langle c_{\tilde i\alpha_{2}}c_{\tilde j\alpha_{4}}\rangle\big\} 
\end{split}
\end{equation}
entering the BdG Hamiltonian. We calculate the superconducting gap as 
\begin{equation}
\Delta_{\tilde j\tilde i\alpha_{2}\alpha_{4}} = \displaystyle\sum_{m=0}^{3}d_{m\tilde j\tilde i}\big(\sigma_{m}i\sigma_{2}\big)_{\alpha_{2}\alpha_{4}}
\label{eq:delta_definition_with_d}
\end{equation}
where $d_{0\tilde j\tilde i}$ is the singlet component and $d_{m\tilde j\tilde i},\ m=1,2,3$ are the triplet components
\begin{equation}
\label{eq:dvector}
d_{m\tilde j\tilde i} =  \Gamma_{m\tilde j\tilde i}\displaystyle\sum_{\alpha_{1}\alpha_{3}}\big( \sigma_{m}i\sigma_{2}\big)^{\dagger}_{\alpha_{3}\alpha_{1}}\langle c_{\tilde i\alpha_{1}}c_{\tilde j\alpha_{3}}\rangle\,,
\end{equation}
calculated self-consitently from the eigenvalues and eigenvectors of the Hamiltonian.

\begin{figure}[tb]
    \centering
    \includegraphics[width=1\linewidth]{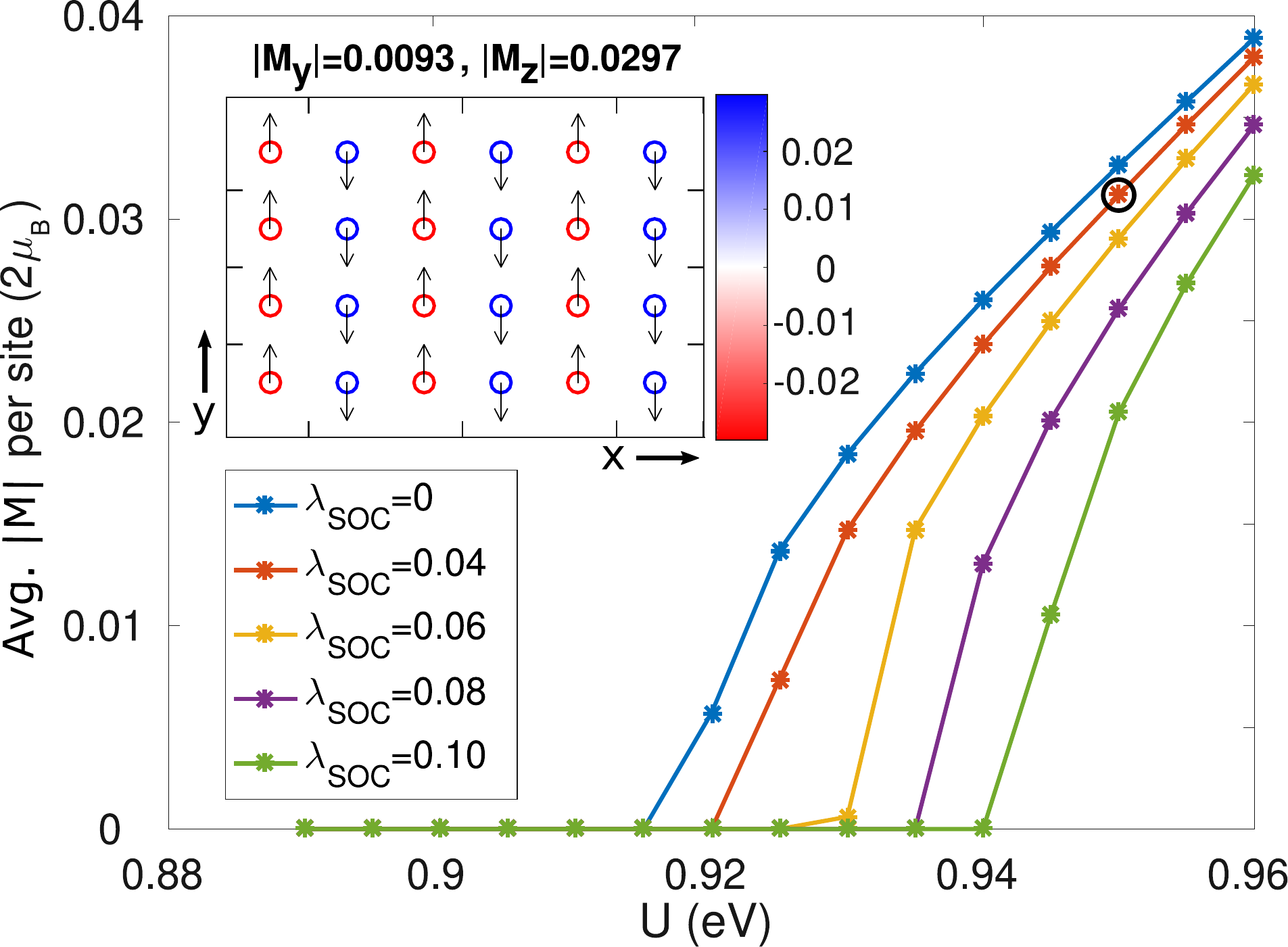}
    \caption{Average absolute magnetization per site as one increases the Hubbard-Hund interaction parameter $U$. Other parameters are related to $U$ via the relation $J=J'=U/4=U'/2$. Temperature was set at $kT = 0.01\ \mathrm{eV}$. Critical interaction parameter increases as the spin-orbit coupling increases. Inset shows the homogeneous system configuration corresponding to the encircled data point. The color scale represents $M_{z}$ and the arrows represent ${\mathbf M}_{xy}$.
    } \label{fig:U_vs_absM_phase_diag_normal_state_v2_inkscape_ylabel_shorter}
\end{figure}

\subsection{Spin-orbit Interaction}
Finally, an atomic spin-orbit coupling is introduced as 
\begin{equation}
H_{\mathrm{SOC}} = \lambda_{\mathrm{SOC}}\ {\mathbf{L}\cdot\mathbf{S}}
\end{equation}
where $\lambda_{\mathrm{SOC}}$ is the spin-orbit coupling parameter for Fe-$3d$ orbitals that can be tuned, $\mathbf{L}$ and $\mathbf{S}$ are respectively the orbital and spin angular momentum operators  evaluated in the orbial basis which can accurately capture the spin-orbit splittings when just onsite terms are included in the Hamiltonian \cite{PhysRevB.88.094522}. We use a spin-orbit coupling of $\lambda_{\mathrm{SOC}} = 0.02\ \mathrm{eV}$ that lies in the typical range for iron-based superconductors \cite{Borisenko2016}.
\begin{figure*}[tb]
    \centering
    \includegraphics[width=1\linewidth]{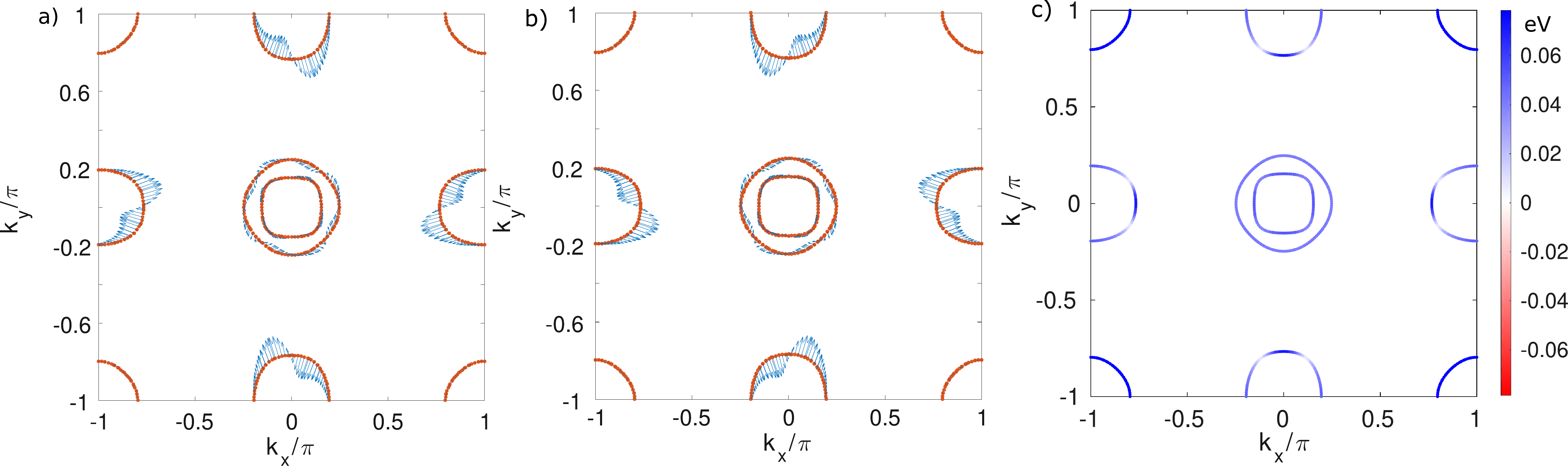}
    \caption[Triplet superconducting gap  induced by spin-orbit coupling 
    ]{Pseudospin triplet superconducting gap induced by spin-orbit coupling. 
    Arrows in a), b) are respectively the real and imaginary parts of the triplet order parameter vector $\vec{d}({\bf k})$ transformed to band space. Only in-plane part $\vec{d_{\parallel}}$ is non-zero. c) shows the full superconducting gap on the Fermi surface. The induced triplet components make the Fermi surface fully gapped. Some points on the Fermi surface may visually appear as nodes, however they have tiny gaps with very faint color. The existence of the triplet component of the order parameter shown is essential for the existence of the nontrivial topological state.} \label{fig:triplet_abs_gap_inkscape_[calc_files_U0_Vimp0_SOC_0.02_for_Fermi_Surf]}
\end{figure*}
Note that we do not include an additional {\it explicit} Rashba coupling that would arise in case of inversion symmetry breaking present at the surface of a bulk material.  We will see below that a similar term is generated by the local magnetic state near the defect.

\section{Results}
\subsection{Homogeneous Normal State Characterization}
We start the discussion of our model with a self-consistent calculation of the chemical potential $\mu_0$ of the Hamiltonian
\begin{equation}
\begin{split}
H &= H_{0} + H_{\mathrm{SOC}} \\&= {\displaystyle \sum_{ij\mu\nu\sigma}\big(t^{\mu\nu}_{ij}c_{i\mu\sigma} - \delta_{ij}\delta_{\mu\nu}\mu_{0}\big)c^{\dagger}_{i\mu\sigma}c_{j\nu\sigma}}
+ 
\lambda_{\mathrm{SOC}}\ {\mathbf{L}\cdot\mathbf{S}}\,,\label{eq:H_normal}
\end{split}
\end{equation}
such that the filling { is fixed at} $n=5.95$ electrons per lattice point, calculated at a temperature of $T=10\;\mathrm{K}$.
The corresponding bands and Fermi surface (as calculated from simply transforming the tight binding model to momentum space) are presented in Fig. \ref{fig:combined_orbital_weights_calc_files_U0_Vimp0_SOC_0pt02_for_Fermi_Surf_bandstructure_inkscape}.

Next, we add the mean field description of the Hubbard-Kanamori Hamiltonian, Eq.~(\ref{eq:H_HbHu_MF}), initialize with random expectation values and calculate the magnetization self-consistently to explore the magnetic phase diagram and by varying the overall magnitude of the interaction parameters to determine the critical values of the Hubbard-Kanamori interactions $U,U^{\prime},J,J^{\prime}$.  As presented in Fig.~\ref{fig:U_vs_absM_phase_diag_normal_state_v2_inkscape_ylabel_shorter}, the magnetic instability of stripe magnetism requires  increasingly high critical $U$ as the spin orbital coupling $\lambda_{\mathrm{SOC}}$ is increased, and the magnetization direction gets locked in the  y-z plane with predominantly z-component (see inset of the Figure).

\subsection{Homogeneous Spin-orbit Coupled Superconducting State}

Next we study the superconducting properties of the multiband system in presence of spin orbit coupling. Assuming in the simplest case only a pairing interaction in the singlet channel, we employ the Hamiltonian
\begin{equation}
    H = H_{0}  + H_{\mathrm{SC}}^{\mathrm{MF}} + H_{\mathrm{SOC}}
\end{equation}
It is to be noted that {\it a priori} the superconducting order parameter in $H_{\mathrm{SC}}^{\mathrm{MF}}$ is not known and is calculated from the eigenvalues and eigenstates of the BdG Hamiltonian via self-consistent iterations.
We choose the superconducting pairing coefficients $\Gamma$  on next-nearest neighbor bonds to be
$0.5\;\mathrm{eV}$, yielding a $s_\pm$ groundstate as known to exist in Fe-based superconductors,  corresponding to a minimum spectral gap in the homogeneous system of $\approx 23\ \rm{meV}$. 

The order parameter is only of spin singlet nature in orbital space, but the multiband nature  introduces  a {subdominant} spin-triplet component in the superconducting gap when expressed in band space by use of the unitary transformation that diagonalizes the normal state Hamiltonian, Eq.~(\ref{eq:H_normal}) in presence of spin-orbit coupling. For details, we refer to appendix (\ref{SingleInducedTriplet}).

In the homogeneous system, we may Fourier transform Eq.~(\ref{eq:delta_definition_with_d}) to
obtain $\underline{\Delta}({\bf k}) =( \underline{d}_0({\bf{k}}) + \underline{\vec d}({\bf k})\cdot \vec\sigma)i\sigma_y$, where underlined quantities are matrices in orbital space.
In  Fig.~\ref{fig:triplet_abs_gap_inkscape_[calc_files_U0_Vimp0_SOC_0.02_for_Fermi_Surf]}, we present the Fermi surface again as contour with arrows describing the (a) real part and (b) imaginary part of the vector $\vec{d}({\bf k})=\vec{d}_{\parallel}({\bf k})+d_z({\bf k}) \vec e_z$ (see Eq.~\ref{eq:dvector}) for the ``induced" triplet superconductivity which lies in the $x$-$y$ plane, i.e. $d_z=0$.
As shown in panel (c), the superconducting state is fully gapped since, despite the fact that for these parameters the $s_\pm$ component is nodal, either the real or imaginary part of the triplet order parameter is nonzero.

\subsection{Linear Defect and Locally Magnetic State}

\begin{figure}
    \centering
    \includegraphics[width=1\linewidth]{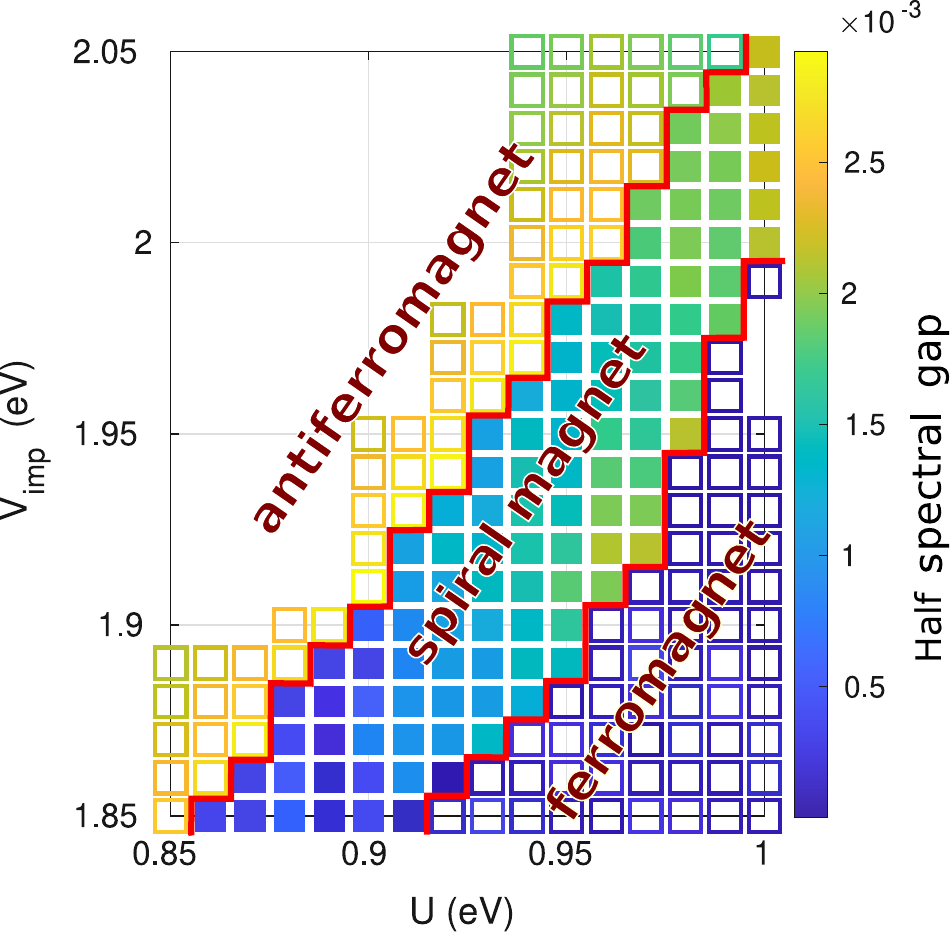}
    \caption[Phase diagram of the system with a vertical defect in presence of the
    (atomic) spin-orbit coupling]{Phase diagram of the system with a vertical defect in presence of the {(atomic)} spin-orbit coupling. The color scale represents half the magnitude of the spectral
    gap. Filled squares, indicate a topological state as characterized by $Q=-1$ and open squares indicate a topologically trivial state as characterized by $Q=+1$.
    `spiral magnet' indicates a configuration where $\mathbf{M}_i$ rotates on the defect axis as shown in Fig.~\ref{fig:spin_spiral_3D_v6_defect_axis_and_tilted} and coincides with the topological region enclosed between the red borders.
    }
    \label{fig:extended_phase_plot_1_v1}
\end{figure}

Finally, we introduce a line of defects as presented in Fig.~\ref{fig:system_with_impurity} and recalculate the mean fields in presence of Hubbard-Hund interactions and superconducting pairing interaction.
Note that in our calculations, we start with a linear defect aligned along the direction towards the nearest neighbor Fe atoms and not along the directions of the Se/Te lattice~\cite{Chen2020}.
In contrast to the work in Ref.~\cite{PhysRevLett.128.016402}, we choose the value of the Hubbard parameter $U$ such that spin fluctuations are strong and eventually the Stoner instability will be crossed locally close to the impurity sites, but not for the full system
since empirically no magnetism has been reported in this system.
We scan the $U$ vs $V_{\rm imp}$ parameter space in search of topological and local magnetic states of the system, the result of which is presented in Fig.~\ref{fig:extended_phase_plot_1_v1}.

\begin{figure*}[tb]
    \centering
   \includegraphics[width=1\linewidth]{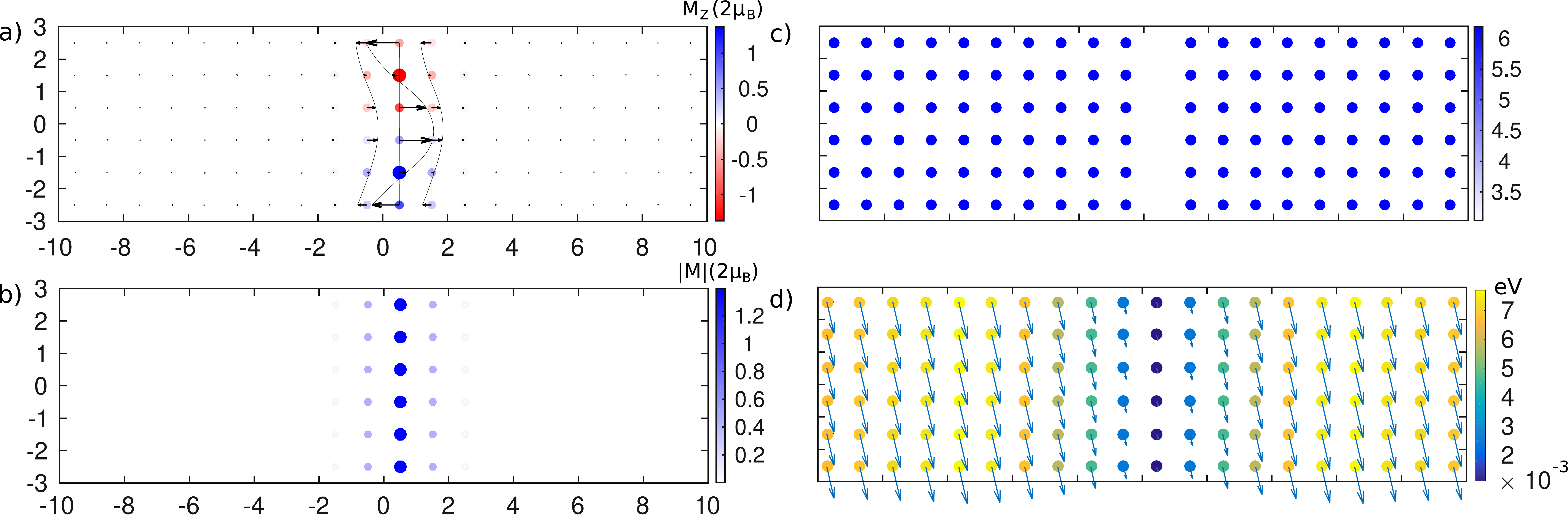}
    \caption{Spin spiral order and superconductivity  (for  $U=0.94\ \mathrm{eV}$, $V_{\mathrm{imp}}=1.9\ \mathrm{eV}$) leading to topologically nontrivial state:
    (a) Visualization of the magnetization vector $\textbf{M}_i$  showing the spiral nature (b) absolute value of the magnetization $\textbf{M}_i$ in the whole system which is only sizeable along the defect line and in the neighborhood (c) local depression in the electron density in the vicinity of the defect and slight enhancement of the same away from defect. (d) Visualization of the superconducting order parameter (magnitude by colorscale) and phase by direction of the  arrows. The order parameter $\bar\Delta_i\equiv \sum_\nu \sum_{j\in NNN(i)} \Delta_{ {\tilde i} {\tilde j}}$ exhibits local depression close to the impurity chain and
    slight modulation (with maxima around $x=\pm 5$ of the same in the entire system.
    {}
    }
    \label{fig:mag_absmag_density_SC_BdG_20x6_gapmap_(x+y)_size_20x6_g0_only_NN2_0pt5.mat_SOC_0.02_U_0.94_VimpStr_str_1.9_inkscape}
\end{figure*}

In the vicinity of,  but slightly below the critical Hubbard repulsion (for driving the full system through the magnetic transition), a sufficiently strong impurity potential gives rise to local magnetic states of the itinerant electrons.
We stress at this point that we have used an entirely spin-rotationally invariant BdG framework to incorporate the interplay of magnetism and superconductivity in channels of differing parity. Specifically, we allow for the formation of noncollinear magnetic order, which can generate an effective spin-orbit interaction when performing a position-dependent rotation of the basis in spin space~\cite{PhysRevB.94.144509,Mascot2021,PhysRevB.104.214501}.
We further notice that the direction of the magnetization will be locked due to the spin orbit coupling, but the phase of the superconducting order parameter comes out to exhibit a random overall phase since for the initial guess a complex order parameter was used.
First, we observe the different configurations of magnetic order on the impurity chain which are labeled in the phase diagram. In the antiferromagnetic phase at large $V_{\rm imp}$, there are alternating magnetic moments as moving along the impurity chain. Working with a small $V_{\rm imp}$ at any particular value of $U$ favors a ferromagnetic chain.
Between these two regimes, the system stabilizes into a spin-spiral magnetic state with the pitch compatible with the boundary condition along the defect chain.  An example of such a spiral magnetic state is presented in Fig. \ref{fig:mag_absmag_density_SC_BdG_20x6_gapmap_(x+y)_size_20x6_g0_only_NN2_0pt5.mat_SOC_0.02_U_0.94_VimpStr_str_1.9_inkscape}. 

\begin{figure}[tb]
    \centering
    \includegraphics[width=1\linewidth]{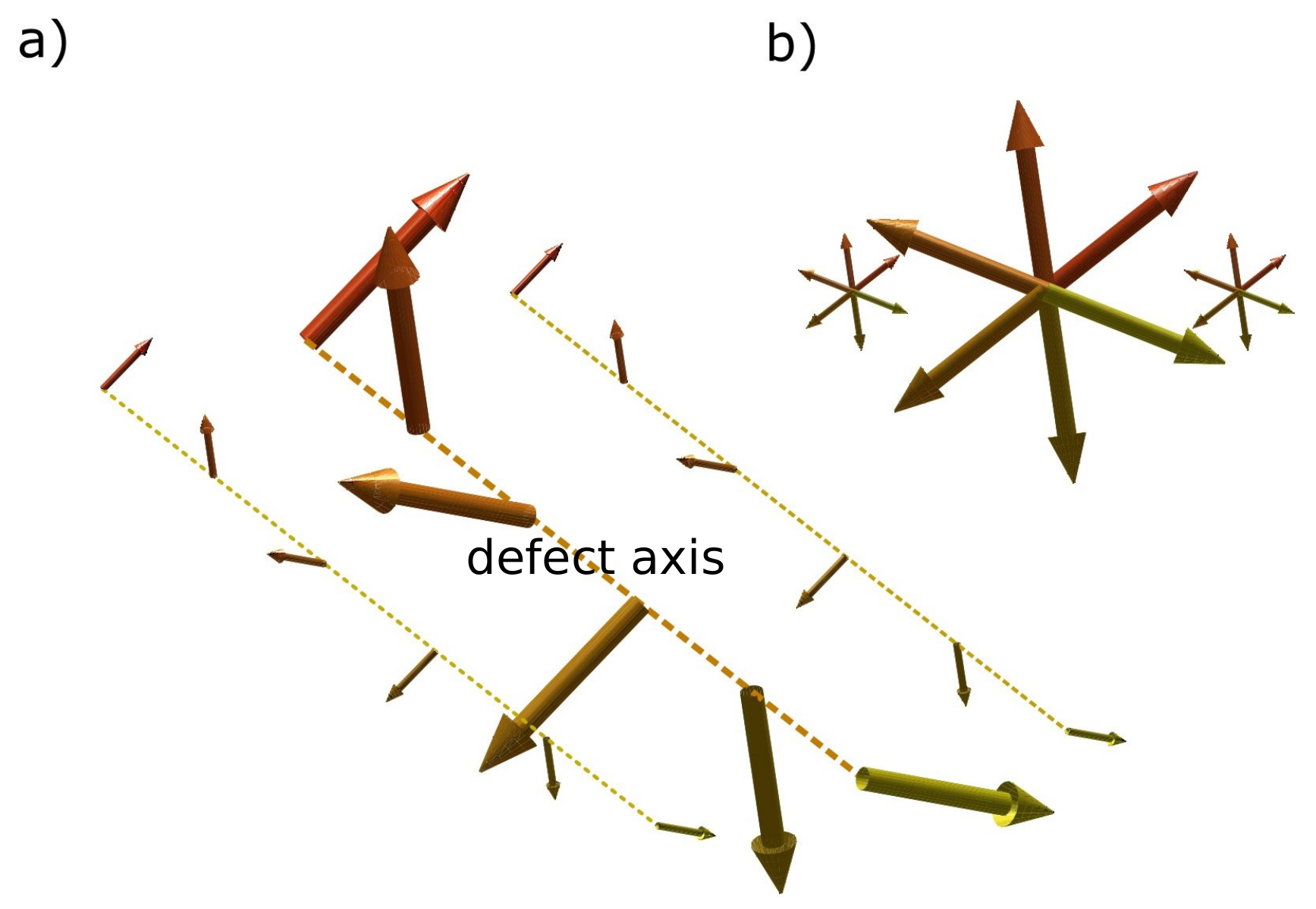}
    \caption[Another visualization of emergent spin-spiral]{{Another visualization of emergent spin-spiral} along the potential impurity defect and in its vicinity: (a) a tilted view (b) view along the defect axis. The variation in color of the arrow {} with angle/rotation is to facilitate visualization and indicates the y-component of the position of ${\mathbf M}_i$.
    }
    \label{fig:spin_spiral_3D_v6_defect_axis_and_tilted}
\end{figure}

Non-zero components of the spiral lie completely in the $x-z$ plane (which is perpendicular to the defect axis). The absolute value of the spiral magnetization is largest on the defect chain but decays rapidly in the direction perpendicular to the chain, see panel (a) of Fig.~\ref{fig:mag_absmag_density_SC_BdG_20x6_gapmap_(x+y)_size_20x6_g0_only_NN2_0pt5.mat_SOC_0.02_U_0.94_VimpStr_str_1.9_inkscape}.
The same can also be seen in Fig.~\ref{fig:spin_spiral_3D_v6_defect_axis_and_tilted} where the induced magnetization for a typical pair of $U$ and $V_{\mathrm{imp}}$ values is shown both in tilted and defect axis view for better visualization.

Indeed, the full system will not be  magnetized in this regime, but the local electron density is suppressed by the impurity potential on the impurity line. As expected from the potential scattering and the additional scattering on the magnetic moment, the superconducting order parameter gets depleted in the vicinity of the defect as expected.

 The topological effect of assumed spin-spiral order in linear magnetic adatom chains has been discussed in the literature earlier \cite{PhysRevB.90.085124, PhysRevB.93.140503, PhysRevB.94.144509}.  However, we report here via our self-consistent microscopic simulations the emergence of stable
  spin-spiral order in an {\it itinerant} electronic system and no magnetic impurities in the vicinity of linear nonmagnetic defects.

\subsection{Linear Defect and Topological State}

Once magnetism develops (in the superconducting state), the Hamiltonian only exhibits particle-hole symmetry and therefore {belongs} to the topological class D of Cartan classification  in one dimension  (see Appendix \ref{LatSymOps}) as shown in table \ref{tab:CartanClassTable}, {thus exhibiting a $Z_2$ invariant in one dimension which is given in Eq. \ref{eq:topoinvPf}.} The (particle) parity operator $P$ has the eigenvalues $(-1)^{\hat N}$ where $\hat N$ is the particle number operator and it commutes with the Hamiltonian, $[H,P]=0$ as given in Eq.~(\ref{eq:all}). Therefore a common set of eigenstates can be found with states having either even or odd particle number (parity).

The $Z_2$ invariant of an infinite (one dimensional) system can be calculated as the product of the parity at the time-reversal invariant momenta (TRIM). Here, we refer to the momenta as obtained from building an infinite one dimensional system by repeating the unit as shown in Fig.~\ref{fig:system_with_impurity} in the $y$ direction and constructing a supercell Hamiltonian $H(k_y)$ as outlined in Appendix \ref{supercell_method}.
In summary, the topological invariant $Q$ for fully gapped Hamiltonians can be calculated as~\cite{PhysRevResearch.3.033049}
\begin{equation}
Q = \displaystyle \prod_{k_{y} \in \rm TRIM} Q(k_y), \ \ \  Q(k_y)=\mathrm{sgn}(\mathrm{Pf}(H(k_y)\tau_x))    
\label{eq:topoinvPf}
\end{equation}
where $\tau_x$ is the  $x$-component of the Pauli matrices in particle-hole space, `Pf' is the Pfaffian (which satisfies for an even-sized skew-symmetric matrix $\mathrm{Pf}(A)^2 = \mathrm{det}(A)$) and the product is taken over time reversal invariant momenta (TRIM), $k_y = 0$ and $k_y = \pi$ in this scenario. We calculate the Pfaffians numerically using a numerical algorithm as described in Ref.~\cite{10.1145/2331130.2331138}.

\begin{figure*}[tb]
    \centering
    \includegraphics[width=1\linewidth]{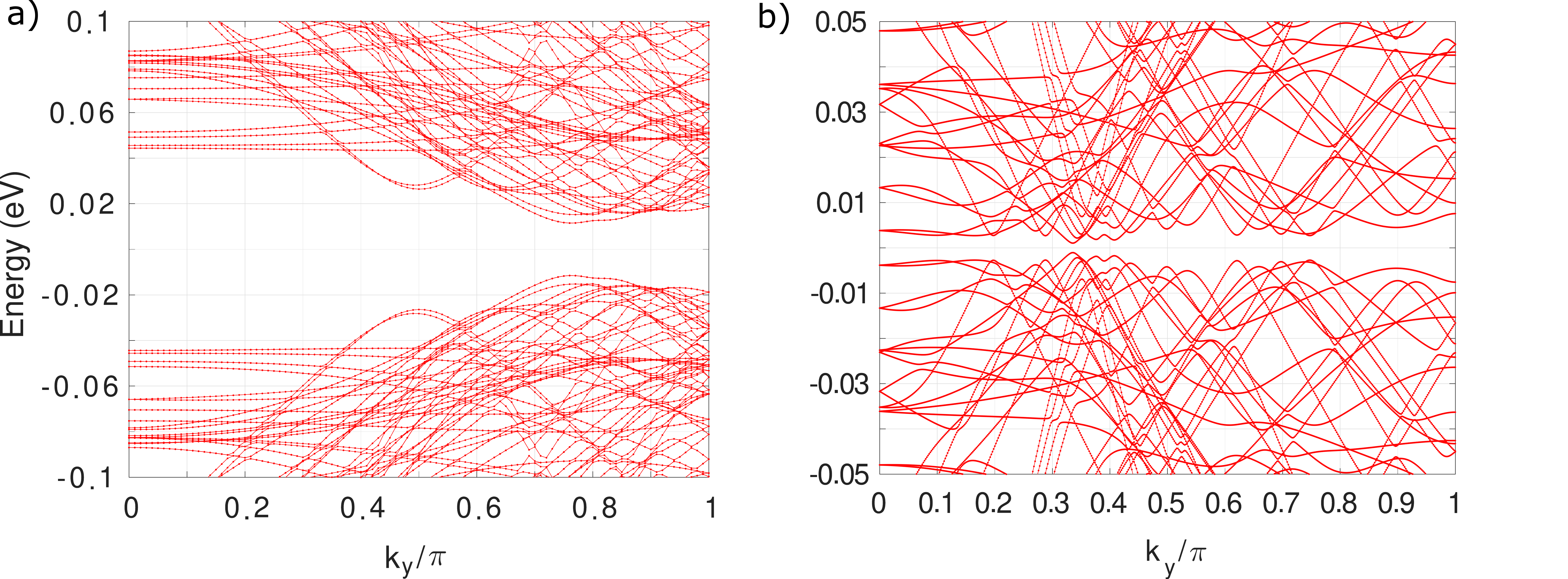}
    \caption{Typical { spectrum} (for   $U=0.925\ \rm{eV}$, $V_{\mathrm{imp}}=1.9\ \rm{eV}$) as a function of the supercell momentum in the system: (a) in absence of the defect, i.e. in homogeneous system {(trivial state)}. 
    (b) {spectrum} in presence of the defect
   {(topological state).}
   } \label{fig:sup_cell_spectr_500_gap_ky_zoomed[size_20x6_g0_NNN_only_SOC_0.02_U_0.925_VimpStr_1pt9]_without_and_with_Vimp}
\end{figure*}

At the same time, we check for the spectral gap by explicit calculation of the eigenvalues of $H(k_y)$ and taking its minimum in the Brillouin zone. Together with the topological invariant, this allows us to complete the topological phase diagram in Fig.~\ref{fig:extended_phase_plot_1_v1} where for each pair of $(U,V_{\mathrm{imp}})$, a full symbol represents $Q=-1$ a topological nontrivial state and an open symbol represents a topologically trivial state with $Q=+1$. The color gives the lowest eigenvalue, i.e. {half} the topological gap for the case $Q=-1$. In Fig.~\ref{fig:sup_cell_spectr_500_gap_ky_zoomed[size_20x6_g0_NNN_only_SOC_0.02_U_0.925_VimpStr_1pt9]_without_and_with_Vimp} we show (a) the typical spectrum of the homogeneous case (where bands have just been folded into the first Briouin zone of the system $6\times 20$ and in panel, (b) the bands in presence of a defect line driving the system into the topological state.

Examining the topological phase diagram, we see that the spiral states
turn out to be topological, with the topological gaps reaching values up to {$0.183\ \Omega_0$ where $\Omega_0\approx 23\ \rm{meV}$} is the spectral gap of the homogeneous system which we for calculation purposes have chosen artificially large. However, our conclusions are expected to be robust since no Lifshitz transition is present on the energy scale of $\Delta_0$ as we used it in our calculations. 

The range of $V_{\mathrm{imp}}$ in which the system supports spiral magnetism shifts up with increasing $U$, as expected. A homogeneous $20\ \times\ 6$ system shows vertical spin stripes above the critical Hubbard-Hund repulsion and the magnetic stability of the stripe increases with increasing repulsion parameter(s). It is therefore expected that higher values of $V_{\mathrm{imp}}$ will be required to twist a vertical ferromagnetic spin-order on the impurity line into an anti-ferromagnetic spin-order through an intermediate spiral-spin-order. To figure out the importance of the spin-orbit coupling, we studied the phase diagram in absence of the atomic spin-orbit coupling as well. Removing the atomic spin-orbit coupling shrinks the area of topological/spiral phase space, although an effective Rashba-type spin-orbit-coupling effect along the defect still remains because of the formation of the spin-spiral order \cite{PhysRevB.90.085124}. Details about the form of the spin-orbit coupling is presented in the Appendix (\ref{EffetiveSOC}). The {effective spin-orbit term generated by the spiral} is not completely identical to the Rashba interaction, but still successfully induces the topological state.

The non-trivial topology of the infinite line defect in a suitable parameter regime has significant implications. Real systems can only have {arrangements of} potential impurities of finite length. As a result, there should be a change in topology from one dimensional defect to the two dimensional bulk at the end of the chain. As this defect subsystem that hosts the non-trivial topology belongs to the class $D$ in one dimension, it should admit a $\mathbb{Z}_2$ invariant, as also corroborated by the two values of the topological invariant $Q$.
Therefore Majorana zero modes are expected to appear at the endpoints of the finite linear defects as edge states.
{\begin{table}[htbp]
    \caption[Cartan classification of topological symmetries of Hamiltonians: The current effective system is one-dimensional and belongs to class D]{Cartan classification of topological symmetries of Hamiltonians: The current effective system is one-dimensional and belongs to class D, shown in blue.{}}\label{tab:CartanClassTable}
    \begin{center}
    \begin{tabular}{|c|c|c|c|c|}
      \hline
      Classes & TRS & PHS & SLS & d=1                       \\
      \hline
      A    & \ 0  & \ 0  & \ 0  &                           \\
      \hline
      AI   & \ 1  & \ 0  & \ 0  &                           \\
      \hline
      AII  &  -1  & \ 0  & \ 0  &                           \\
      \hline
      AIII & \ 0  & \ 0  & \ 1  & \ $\mathrm{\mathbb{Z}}$   \\
      \hline
      BDI  & \ 1  & \ 1  & \ 1  & \ $\mathrm{\mathbb{Z}}$   \\
      \hline
      CII  &  -1  &  -1  & \ 1  & \ $\mathrm{\mathbb{Z}}$   \\
      \hline
      \textcolor{blue}{\textbf{D}} & \ \textbf{\textcolor{blue}0}  & \ \textbf{\textcolor{blue}1}  & \ \textbf{\textcolor{blue}0}  & \ {\color{blue}$\mathrm{\mathbb{Z}_2}$} \\
      \hline
      C    & \ 0  &  -1  & \ 0  &                           \\
      \hline
      DIII &  -1  & \ 1  & \ 1  & \ $\mathrm{\mathbb{Z}_2}$ \\
      \hline
      CI   & \ 1  &  -1  & \ 1  &                           \\
      \hline
    \end{tabular}
    \end{center}
\end{table}}
\section{Discussion and Conclusion}
In summary, we have explored the possibility to realize a topological superconductor on a one dimensional chain of impurity atoms where magnetism is generated from correlations in the multiband electronic structure. For self-consistently calculating the magnetic oder parameter and the (inhomogeneous) superconducting order parameter, we used a spin-rotationally invariant mean field approach to decouple the Hubbard-Hund interaction. The topologically non-trivial state can be reached once a spin-spiral state along the impurity chain forms in vicinity of competing magnetic ground states. The interaction parameters are tuned to mimic physical properties of the Fe(Se,Te) system. {It is known that FeTe is a $(\pi/2,\pi/2)$ double stripe magnet, and that the doped system has strong stripelike $(\pi,0)$ magnetic
correlations\cite{Lietal2021,Baoetal,Shiliangetal,Kozetal}}.  Therefore it is natural to expect Fe(Se,Te) to be in proximity of the spiral  magnetic instability.

The mechanism for topological superconductivity is based on an effectively (pseudo-spin) triplet order parameter forming from the multiband superconductivity in conjunction with the non-collinear magnetic order. Additional (atomic) spin orbit coupling enhances this effect and allows to reach the topolgical phase for a broader set of model parameters.

{One may ask if the type of defect considered here, namely a line of Fe vacancies, is of general relevance to Fe-based systems, and to Fe(Se,Te) in particular, i.e. is it the only type of defect that produces the magnetically driven  topological state?  The observation that motivated our work\cite{doi:10.1126/science.aaw8419,Chen2020} corresponds to a line of 
Se/Te vacancies along a  diagonal direction in the coordinates of the maps shown here.  
In the current formalism, where the electronic structure has been mapped onto a Fe-only tight binding model, this would correspond to a line of potentials on the Fe sites that are neighboring the vacancies, i.e. a double chain.  
Furthermore, one has to consider supercells both in x and y directions for obtaining an Hamiltonian describing an infinite chain along the diagonals in order to calculate the topological invariant from the product of Pfaffians at the TRIM. An alternative approach would be to set up a diamond geometry rather than a rectangle. Another technical issue that we anticipate is the larger unit cell needed because of the double line of impurities, such that we postpone discussion of this case to a future study.}

{We emphasize other novel aspects} of the work described in this paper compared to the previous works in similar directions. Firstly, here we consider a realistic band-structure suitable for iron-based superconducting materials rather than a model single-band structure \cite{PhysRevB.93.140503,PhysRevB.94.144509}.
Secondly and most importantly, all previous works that have dealt with topological consequences of spin chains, whether ferromagnetic or antiferromagnetic or spiral, {assumed  the existence of the chain of localized magnetic moments,} whereas the moment formation described in this paper arises completely self-consistently from itinerant electrons present in the system itself.
We believe that to understand how to recreate and possibly engineer Majorana modes on other platforms, it is important to have a realistic theory
of topological effects arising self-consistently from itinerant electrons and inherent defects in the Fe(Se,Te) system. 

It is important to check possible consistency of our scenario with the observations of experiment, e.g. Ref.  \cite{Chen2020}.  In the experiment, zero-energy bound state peaks at chain ends extending over apparently only a few lattice constants were observed  for defect chains of order $\sim$ 15 lattice constants.  These two facts are consistent with the observed sharp unsplit zero-bias peaks if the chain length is indeed several times the  MZM size.  In results presented in Fig. \ref{fig:extended_phase_plot_1_v1}, the magnitude of the topological spectral gap found in our work can reach a fraction of $0.183\ \Omega_0$ of the spectral gap of the homogeneous superconductor, $\Omega_0$, which is found to be 23 meV in the calculation as stated above.   Since we have -- for numerical purposes -- worked with gap sizes that are artificially large, it seems appropriate to assume that the correct topological gap in our scenario scales with the ratio ($\Omega_{0,{\rm topo}}/\Omega_0$), where $\Omega_0=23$ meV as stated above.   The true gap in FeSe is $\approx 2T_c\approx 30K$, or about 2.6 meV.  So for purposes of our estimate,  we expect a topological gap of roughly $\Delta_{\rm topo}\simeq$ 0.48 meV in Fe(Se,Te) in our scenario.  With a Fermi velocity of Fe(Se,Te) of roughly 0.2 eV-\AA \cite{Feng2010}, this corresponds to a decay length of $\xi\simeq v_F/(\pi \Delta_{\rm topo}) \sim$ 130\AA $\sim$ 33$a$.  This is clearly much larger than the observed bound state radius, and would in addition lead to hybridization of the end chain states and splitting of the zero bias peak, which is not observed.  One therefore needs to ask if, by judicious choice of Hamiltonian parameters,  the topologogical gap can be increased  by nearly an order of magnitude to be of order of the bulk gap. This would then  correspond to a decay length roughly consistent with experiment.

In order to find ways to reach a larger topological gap within our scenario, one first needs to discuss what qualitatively affects its size. In view of the analogy to the Kitaev chain, the gap is determined generally by the overall size of the (pseudospin) triplet component of the order parameter.
In our case, a triplet component can be generated in two ways: The first is discussed in Fig.~\ref{fig:triplet_abs_gap_inkscape_[calc_files_U0_Vimp0_SOC_0.02_for_Fermi_Surf]} where the (atomic) spin orbit coupling induces a triplet component when transforming to band space. Indeed, a larger value of the spin orbit coupling constant $\lambda_{\text{SOC}}$, or considering additionally surface-induced spin orbit coupling (Rashba-type) would increase this effect. It could lead to a triplet component at the Fermi points as large as the singlet order parameter {provided that  spin-flip} terms in the normal state Hamiltonian are dominant at the Fermi surface. The values of the spin orbit coupling are however constrained by experimental measurements \cite{Borisenko2016}, so  that spin-orbit coupling induced band splittings at the Fermi points with gap minima could come close to that limit.  

We have discussed {our result} that even completely without spin-orbit coupling, there is a (small) topological phase. This in {our} view can be understood as follows: To map our Hamiltonian onto an effective Kitaev-like model, one needs to perform local changes of the basis by rotations in spin space in a way that the magnetic moment becomes ferromagnetic everywhere. For a homogeneous superconducting order parameter, {this mapping} will not lead to any (pseudospin) triplet component, while an inhomogeneous order parameter, as obtained in our self consistent mean field approach, transforms to such a component. Therefore, increasing the inhomogeneity in principle increases this component as well. Larger inhomogeneities are expected for (1) increasing the impurity potential $V_{\text{imp}}$ or (2) inducing pairbreaking from magnetic scattering by larger local magnetization, which occurs from increasing $U$. Exactly these tendencies are seen in our phase diagram in Fig.~\ref{fig:extended_phase_plot_1_v1} where larger topological gaps are observed on the right boundary of the topological phase (large $U$ and therefore larger ${\mathbf M}_i$), i.e. {in the direction of} larger impurity potentials. In our self-consistent approach this can however not be tuned into a sweet spot because larger $U$ will turn the local magnetization into a ferromagnet and  large $V_{\text{imp}}$ closer to unitary scatterers makes the width (in $U$) of the phase with spiral magnetism smaller.  

Other physical parameters in the specific case we have evaluated here may be  less than optimal for the creation of a robust, localized topological state.  For example, we have not explored the role of varying Hund's exchange $J$ ratio to Coulomb strength $U$, etc.  Numerical cost constraints have generally prevented us from exploring parameter space fully.  While we intend to follow up the most promising avenues, some of which we have identified here, our principal result and conclusion is that topological 
states can indeed be driven by self-organized spiral states around nonmagnetic impurity chains, and that Rashba spin-orbit coupling due to the surface, while it may play a role, is not necessary for the effect.  Whether or not the end-chain bound states observed in experiment are indeed isolated Majorana zero modes remains an open question, but a plausible mechanism to create them indeed exists. 
\vskip .2cm
{\it Acknowledgements} We thank B.M. Andersen and M. Christensen for helpful discussions.  M. P. and P.J. H. were supported by  NSF-DMR-1849751.  A.K. acknowledges support by the Danish National Committee for Research Infrastructure (NUFI) through the ESS-Lighthouse Q-MAT.
\appendix
\section{Supercell Method}
\label{supercell_method}
\begin{figure}[tb]
    \centering
    \includegraphics[width=1\linewidth]{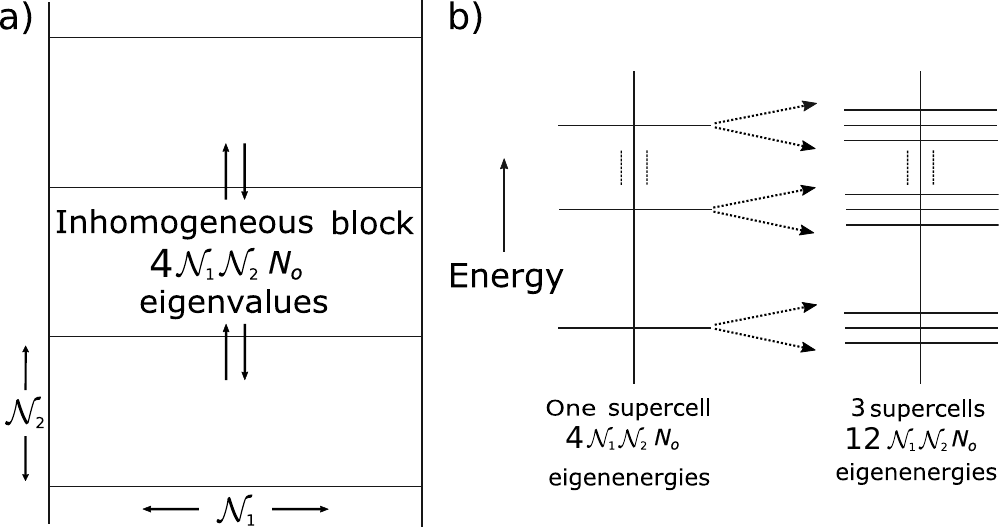}
    \caption[Illustration of the supercell method]{{Illustration of the supercell method:} A supercell lattice of size $3$ consisting of supercells of size $\mathcal{N}_{1} \times \mathcal{N}_{2}$ (periodically continued vertically) where each lattice site hosts $N_{o}$ orbitals. The double-way arrows show the supercell-supercell coupling in which the sites near the supercell boundaries participate. {}}
    \label{fig:supercell}
\end{figure}

{Due to translation invariance, real space BdG calculations can instead be readily obtained from the corresponding equations in momentum space in case of a homogeneous system.}
Arbitrarily dense momentum space can be chosen for this, which in real space would correspond to a bigger and bigger part of an infinite homogeneous lattice. One can also construct this `larger portion of lattice' by stitching the `lattice-portion corresponding to the coarse momentum grid' together one after another.
{}
In the current work, we only need this supercell construction in the vertical direction (Fig. \ref{fig:supercell}, panel a)). The `entire lattice-portion corresponding to the coarse momentum grid' is usually named a `supercell' in real space. The momentum corresponding to the periodicity of these supercells is termed `supercell momentum'. Obviously, the neighboring supercells couple with each other through the sites near their margin/boundary only, the supercell being just an imaginary construct of grouping lattice sites together. Hence, for a rectangular lattice (which has been used in this article), only nearest neighbor (NN) can couple with each other unless the supercell size is not critically small compared to the hopping range of the original lattice sites. 
  
The corresponding scenario of stitching together the inhomogeneous lattices of a finite size is particularly useful for increasing smoothness of the spectrum. In this case, 
{within the supercell there is no notion of momentum because of lack of translational invariance and} 
the eigenstates of the inhomogeneous finite lattice are to be obtained by numerical diagonalization, often starting from the real space basis. However, subsequently, the use of many supercells effectively creates a narrow band of energy around each of these numerically obtained eigenenergies (Fig. \ref{fig:supercell}, panel b)), thereby increasing the spectrum resolution. 

If $M$ supercells are used, the supercell BdG equation in matrix form can be written as
\begin{equation}
    \begin{pmatrix}
    H({k_y}) & \Delta({k_y}) \\
    \Delta^{\dagger}({k_y}) & -H^T(-{k_y})
    \end{pmatrix}
    U_{:,ns}({k_y})
    =
    E_{ns}({k_y})
        U_{:,ns}({k_y})
    \label{A1}
\end{equation}
with 
\begin{align}
  H({k_y}) = \sum_{I} T_{I0} e^{i{k_y}{Y}_I}, \label{A2}
  \\
  \Delta({k_y}) = \sum_{I} \Delta_{I0} e^{i{k_y}{Y}_I}.
\end{align}
Here ${k_y}$ is supercell momentum, ${Y}_{I}$ is the coordinate of the $I$th supercell, $T_{I0}$ is the supercell hopping matrix (of size $2\mathcal{N}_1 \mathcal{N}_2 N_o \times 2\mathcal{N}_1 \mathcal{N}_2 N_o$ where the size of the supercell is $\mathcal{N}_1 \times \mathcal{N}_2$ and each site has $N_o$ orbitals with two possible spins) between the supercell at origin and the $I$th supercell, and  $\Delta_{I0}$ is the supercell gap matrix (of size $2\mathcal{N}_1\mathcal{N}_2N_o \times 2\mathcal{N}_1\mathcal{N}_2N_o$ containing both the singlet and triplet components) {}between the supercell at origin and the $I$th supercell.
The matrix $U$ diagonalizes the Hamiltonian and $ U_{:,ns}({k_y})$ is the $ns$ column-vector of $U({k_y})$.
In this work we have $N_o=5$ and number of lattice points $N=\mathcal{N}_1\mathcal{N}_2$. Most of the matrix elements of $T_{I0}$ and $\Delta_{I0}$ (for $I\ne 0$) are zero unless they involve sites near the supercell-boundary which maintain the supercell-supercell coupling. $T_{00}$ and $\Delta_{00}$ are exactly same as the hopping matrix and the gap matrix {(excluding periodic terms)} of the $\mathcal{N}_1 \times \mathcal{N}_2$ inhomogeneous lattice, solved by explicit iterative self-consistent diagonalization to get eigenenergies indexed by $n$. $k_y$ can take values as $k_y=\dfrac{m}{M},\ m=0,1,2,...M-1$ in units of $\dfrac{2\pi}{\mathcal{N}_{2}a}$ where $a$ is lattice constant, and for each  ${k_y}$ the supercell BdG diagonalization is done just once to obtain the eigenvectors $U_{:,ns}({k_y})$, each of size $4\mathcal{N}_1\mathcal{N}_2N_o$. So, as already mentioned, corresponding to each $n$, a band of additional $M - 1$ eigenenergies are obtained. The entire scheme is presented in Fig.~\ref{fig:supercell}.
\section{Matrix Representation of Symmetries in the System for Topological Classification}
\label{LatSymOps}
The fermionic basis used in the particle-hole and spin-space is 
\begin{eqnarray}
c^{\dagger} = 
\begin{pmatrix}
c_{\uparrow}^{\dagger} & 
c_{\downarrow}^{\dagger} &
c_{\uparrow} & 
c_{\downarrow}
\end{pmatrix}.
\end{eqnarray}
Under the particle-hole transformation this goes to 
\begin{eqnarray}
c^{\dagger} = 
\begin{pmatrix}
c_{\uparrow}^{\dagger} & 
c_{\downarrow}^{\dagger} &
c_{\uparrow} & 
c_{\downarrow}
\end{pmatrix}
\rightarrow
\begin{pmatrix}
c_{\uparrow} & 
c_{\downarrow} &
c_{\uparrow}^{\dagger} & 
c_{\downarrow}^{\dagger}
\end{pmatrix}.
\end{eqnarray}
So, the particle hole transformation operator in the particle-hole and spin space reads
\begin{eqnarray}
\begin{pmatrix}
c_{\uparrow}^{\dagger} \\
c_{\downarrow}^{\dagger} \\
c_{\uparrow} \\ 
c_{\downarrow}
\end{pmatrix}
= 
\underbrace{
\begin{pmatrix}
0 & 0 & 1 & 0 \\
0 & 0 & 0 & 1 \\
1 & 0 & 0 & 0 \\
0 & 1 & 0 & 0
\end{pmatrix}}_{M_\mathrm{P}^\mathrm{PH,spin} =\ \sigma_{x}^{\mathrm{PH}} \otimes\ \mathbb{I}_2^{\mathrm{spin}}}
\begin{pmatrix}
c_{\uparrow} \\ 
c_{\downarrow} \\
c_{\uparrow}^{\dagger} \\ 
c_{\downarrow}^{\dagger}
\end{pmatrix}
\end{eqnarray}
The full particle-hole transformation operator reads
\begin{eqnarray}
M_{\mathrm{P}} =\ \sigma_{x}^{\mathrm{PH}} \otimes \mathbb{I}_2^{\mathrm{spin}}
\otimes \ \mathbb{I}_{MN}^{\mathrm{lattice}} \otimes  \mathbb{I}_5^{\mathrm{orbital}}
\end{eqnarray}
where the system size is $M \times N$.

Under the time-reversal transformation the basis goes to
\begin{eqnarray}
c^{\dagger} = 
\begin{pmatrix}
c_{\uparrow}^{\dagger} & 
c_{\downarrow}^{\dagger} &
c_{\uparrow} & 
c_{\downarrow}
\end{pmatrix}
\rightarrow
\begin{pmatrix}
i c_{\downarrow}^{\dagger} & 
-i c_{\uparrow}^{\dagger} &
i c_{\downarrow} & 
-i c_{\uparrow}
\end{pmatrix}
\end{eqnarray}

So, the time-reversal transformation operator in the particle-hole and spin space reads

\begin{eqnarray}
\begin{pmatrix}
-ic_{\downarrow} \\
ic_{\uparrow} \\
-ic_{\downarrow}^{\dagger} \\ 
ic_{\uparrow}^{\dagger}
\end{pmatrix}
= 
\underbrace{
\begin{pmatrix}
0 & -i & 0 & 0 \\
i & 0 & 0 & 0 \\
0 & 0 & 0 & -i \\
0 & 0 & i & 0
\end{pmatrix}}_{M_\mathrm{T}^\mathrm{PH,spin} =\ \mathbb{I}_2^{\mathrm{PH}}\otimes \ \sigma_{y}^{\mathrm{spin}}} 
\begin{pmatrix}
c_{\uparrow} \\ 
c_{\downarrow} \\
c_{\uparrow}^{\dagger} \\ 
c_{\downarrow}^{\dagger}
\end{pmatrix}
\end{eqnarray}

The full time-reversal transformation operator reads
\begin{eqnarray}
M_{\mathrm{T}} =\ \mathbb{I}_2^{\mathrm{PH}}\otimes \ \sigma_{y}^{\mathrm{spin}}
\otimes \ \mathbb{I}_{MN}^{\mathrm{lattice}} \otimes  \mathbb{I}_5^{\mathrm{orbital}}
\end{eqnarray}
\section{Triplet Superconductivity from Singlet Component and Spin-orbit Coupling}
\label{SingleInducedTriplet}
Let's denote the unitary matrix that diagonalizes the normal state momentum-space Hamiltonian for spin $\uparrow$ as $U_{\uparrow\uparrow}(\mathbf{k})$ and its counterpart for spin $\downarrow$ as $U_{\downarrow\downarrow}(\mathbf{k})$. Then the transformation matrix for the normal state Hamiltonian in Nambu space reads as

\begin{equation}
\begin{pmatrix}
U_{\uparrow\uparrow}(\mathbf{k}) & 0 & 0 & 0\\
0 & U_{\downarrow\downarrow}(\mathbf{k}) & 0 & 0\\
0 & 0 & U^{*}_{\uparrow\uparrow}(-\mathbf{k}) & 0\\
0 & 0 & 0 & U^{*}_{\downarrow\downarrow}(-\mathbf{k})
\end{pmatrix}
\end{equation}
Hence \textit{in absence of spin-orbit coupling} the superconducting gap matrix reads as
\begin{equation}
\begin{split}
\Delta =
&\begin{pmatrix}
U^{\dagger}_{\uparrow\uparrow}(\mathbf{k}) & 0 \\
0 & U^{\dagger}_{\downarrow\downarrow}(\mathbf{k})
\end{pmatrix}
\begin{pmatrix}
\Delta_{\uparrow\uparrow}(\mathbf{k}) & \Delta_{\uparrow\downarrow}(\mathbf{k}) \\
 \Delta_{\downarrow\uparrow}(\mathbf{k}) & \Delta_{\downarrow\downarrow}(\mathbf{k})
\end{pmatrix}
\\\times&\begin{pmatrix}
U^{*}_{\uparrow\uparrow}(-\mathbf{k}) & 0 \\
0 & U^{*}_{\downarrow\downarrow}(-\mathbf{k})
\end{pmatrix}\\
=
&\begin{pmatrix}
U^{\dagger}_{\uparrow\uparrow}(\mathbf{k}) \Delta_{\uparrow\uparrow}(\mathbf{k}) U^{*}_{\uparrow\uparrow}(-\mathbf{k})
&
U^{\dagger}_{\uparrow\uparrow}(\mathbf{k}) \Delta_{\uparrow\downarrow}(\mathbf{k}) U^{*}_{\downarrow\downarrow}(-\mathbf{k}) 
\\
U^{\dagger}_{\downarrow\downarrow}(\mathbf{k}) \Delta_{\downarrow\uparrow}(\mathbf{k}) U^{*}_{\uparrow\uparrow}(-\mathbf{k})
&
U^{\dagger}_{\downarrow\downarrow}(\mathbf{k}) \Delta_{\downarrow\downarrow}(\mathbf{k}) U^{*}_{\downarrow\downarrow}(-\mathbf{k})
\end{pmatrix}
\end{split}
\end{equation}
in the same  basis. So, clearly there is no mixture among equal-spin and opposite-spin components of the superconducting gaps.

However, when a \textit{spin-orbit coupling} is present, $U_{\uparrow\uparrow}(\mathbf{k})$ and $U_{\downarrow\downarrow}(\mathbf{k})$ are no longer good diagonalizing matrices for the respective spins and the full normal state diagonalizing matrix looses the block-diagonal character by developing small off-diagonal blocks

\begin{equation}
\begin{pmatrix}
U_{\uparrow\uparrow}(\mathbf{k}) & M_{1}(\mathbf{k}) \\
  M_{2}(\mathbf{k}) & U_{\downarrow\downarrow}(\mathbf{k})
\end{pmatrix}    
\end{equation}
so that now the superconducting gap matrix reads as
\begin{equation}
\begin{split}
\Delta \approx
&\begin{pmatrix}
U^{\dagger}_{\uparrow\uparrow}(\mathbf{k}) &  M^{\dagger}_{2}(\mathbf{k}) \\
 M^{\dagger}_{1}(\mathbf{k}) & U^{\dagger}_{\downarrow\downarrow}(\mathbf{k})
\end{pmatrix}
\begin{pmatrix}
\Delta_{\uparrow\uparrow}(\mathbf{k}) & \Delta_{\uparrow\downarrow}(\mathbf{k}) \\
 \Delta_{\downarrow\uparrow}(\mathbf{k}) & \Delta_{\downarrow\downarrow}(\mathbf{k})
\end{pmatrix}\\\times&
\begin{pmatrix}
U^{*}_{\uparrow\uparrow}(-\mathbf{k}) & M^{*}_{1}(-\mathbf{k}) \\
 M^{*}_{2}(-\mathbf{k}) & U^{*}_{\downarrow\downarrow}(-\mathbf{k})
\end{pmatrix} \\
\end{split}
\end{equation}
in the same  basis that diagonalizes {the} full normal state Hamiltonian.
{Hence, clearly the singlet component of the superconducting gap contributes to the pseudo-spin triplet component.}
Obviously this triplet {component} has no equal-spin character. It is to be noted that no spin-triplet superconductivity is explicitly driven in the self-consistency by any non-zero spin-triplet pairing coefficient. Hence there is no dominant $\Delta_{\uparrow\uparrow}(\mathbf{k})$ or $\Delta_{\downarrow\downarrow}(\mathbf{k})$ which could contribute to the pseudo-singlet component because of spin-mixing. This also explains why the $\mathbf{d}$ vector for induced triplet component lies entirely in the $xy$ plane.  

\section{Spin-spiral Induced Effective Spin-orbit Coupling}
\label{EffetiveSOC}
The spin spiral in the self-consistent solution rotates on the $xz$-plane as a function of $y$ coordinate. It can be described as 
\begin{equation}
\mathbf{S} = S(\mathrm{cos}(P y_{i} - \phi),\ 0,\ \mathrm{sin}(P y_{i} - \phi))\,,
\end{equation}
where $P$ is the pitch of the spiral.
A position dependent gauge transformation in the spin-space of the fermion operators as 
\begin{equation}
\begin{pmatrix}
{c}_{i\uparrow} \\
{c}_{i\downarrow}
\end{pmatrix}
\rightarrow
e^{i \sigma_{y} P y_{i}}
\begin{pmatrix}
\tilde{c}_{i\uparrow} \\
\tilde{c}_{i\downarrow}
\end{pmatrix}
\end{equation}
effectively aligns the quantization axis along the local self-consistent spin direction. As a result, the effective hopping between sites $i$ and $j$ reads as
\begin{equation}
\begin{split}
&H_0^{\mathrm{eff}} 
=
\sum_{ij}t_{ij}
\begin{pmatrix}
\tilde{c}^{\dagger}_{i\uparrow} & \tilde{c}^{\dagger}_{i\downarrow}
\end{pmatrix}
e^{-i \sigma_{y} P (y_{i} - y_{j})}
\begin{pmatrix}
\tilde{c}_{i\uparrow} \\
\tilde{c}_{i\downarrow}
\end{pmatrix}
=\\&
\sum_{ij}t_{ij}
\begin{pmatrix}
\tilde{c}^{\dagger}_{i\uparrow} & \tilde{c}^{\dagger}_{i\downarrow}
\end{pmatrix}
\begin{pmatrix}
\mathrm{cos}\{P(y_{i} - y_{j})\} & -\mathrm{sin}\{P(y_{i} - y_{j})\} \\
\mathrm{sin}\{P(y_{i} - y_{j})\} & \mathrm{cos}\{P(y_{i} - y_{j})\}
\end{pmatrix}\\&\ \ \ \ \ \ \ \ \ \ \ \ \ \ \ \ \ \ \ \times
\begin{pmatrix}
\tilde{c}_{j\uparrow} \\
\tilde{c}_{j\downarrow}
\end{pmatrix}
\end{split}
\end{equation}
Thus, the spin-spiral generates an effective spin-orbit coupling term in the effective tight-binding Hamiltonian.
\bibliography{dislocation}
\end{document}